Pomazanov Mikhail

# Assessment of loan losses after default

The paper shows how to determine the loss on an LGD borrower's loan after default, with or without preparation of a separate model. LGD after default is estimated taking into account the average repayment period of the defaulted loan, knowledge of volumes, moments of default and repayments, the rate or other parameters in the vector of determinants. The calculation of the average repayment period for overdue loans is given in the article. A Bayesian scheme is used to estimate repayable debts, considering the percentage of repayment. A general recovery model was used for the LGD segment recovery process. Only this type of model allows you to set LGD less than or equal to 1, which is required for further estimates.

Помазанов М.В.

# Оценка потерь для ссуд после дефолта

В работе показано как определять убыток по кредиту заемщика LGD после дефолта с подготовкой или без подготовки отдельной модели. LGD после дефолта оценивается с учетом среднего срока возврата дефолтной ссуды, знания объёмов, моментов дефолта и погашений, ставки или прочих параметров в векторе детерминантов. Расчет среднего срока возврата просроченных кредитов, приведен в статье. Используется байесовская схема оценки возвращаемых долгов с учетом доли возврата. Использовалась модель общего взыскания на процесс восстановления по сегменту LGD, только такой тип модели позволяет установить $\mathrm{LGD} \leq 1$, что требуется для выкладок дальнейших оценок.

## 1. Введение

Величина убытка по кредиту заемщика в случае дефолта (LGD) является важным параметром для определения ожидаемых и неожиданных потерь кредитного риска. Величина LGD определяется для двух состояний ссуды: без дефолта (LGD WD) и в дефолте (LGD ID). Признание применяемой модели LGD идет на основании поступивших платежей по истории поведения дефолтных ссуд. Модели LGD WD разрабатываются для различных условий ссуд юридическим и физическим лицам, на широком списке факторов и сегментов. Имеется существенный объем публикаций по моделям LGD WD, даны важные установки того, как разрабатывать модель [1-6]. По моделям LGD ID аналитические публикации фактически отсутствуют. Поэтому было решено разработать модель LGD ID на минимальном и доступном количестве факторов, которую можно применять для практики построения LGD для того, кто получил дефолт. LGD ID приближается к 100% на горизонте событий, сопоставимом со средним сроком погашения ссуды. Поэтому само состояние LGD ID отличное от 100% длится не долго и оказывает на величину резервов по всем ссудам не высокое значение, сравнимое с погрешностью оценки главной величины, которая есть LGD WD. Значению LGD ID не оказывалось необходимое внимание, поэтому наша задача определить эту величину и указать компоненты для эффективной оценки. На входе мы имеем: дату дефолта, рассчитанное значение WD, общее поведение ссуд в дефолте, платежи, которые поступили на счет дефолтного заемщика, а также некие решения заемщика по обслуживанию ссуды.

В текущем исследовании мы считаем, что модель LGD внедрена, это может быть модель только до дефолта, но также и модель учета LGD в момент дефолта, но после дефолта. После дефолта LGD будет меняться в зависимости от начисления погашения, скоростных характеристик и прочих детерминантов вплоть до $LGD = 100\%$.

## 2. Учет затрат на взыскание и базовые формулы для LGD

Для заемщика в дефолте из выборки мощности $M$ имеется задолженность EAD на момент дефолта $d$, равная $\hat{E}_0$, а также восстановления после дефолта $RR_n$ (денежных единиц) и стоимость затрат на изъятие дефолтных долгов (издержек) $C_n$ на каждый месяц после дефолта $n = 1, \dots \infty$. Имеется максимальный горизонт взыскания N (месяцев), установленный для рассматриваемого сегменту кредитный требований (КТ) применения модели. Математически LGD WD на момент дефолта $t = 0$ рассчитывается по формуле

$$\widehat{LGD}(0) = 1 - \frac{1}{\hat{E}_0} \sum_{n=1}^{N} \frac{RR_n - C_n}{(1+r)^{\frac{n}{12}}} \tag{1}$$

где $r$ – ставка дисконтирования, мы применяем предположение неизменной ставки после дефолта, например, стоимость кредита. Оценка ставки дисконтирования $r$ не имеет единой фундаментальной основы. При ее выборе следует опираться на один из принципов, избранных банком. Значение $r$ может быть равной ставке межбанковского кредитования, действующей на момент дефолта, плюс 5%. Либо $r$ есть эффективная ставка кредитования, либо полная стоимость кредита на момент дефолта, либо еще другие принципы. Подробнее, разные предложения ставки можно увидеть в списке работ [7-9].

Значение издержек по изъятию платежа на взыскание полагается пропорциональным самим восстановлениям

$$C_n = c \cdot RR_n \tag{2}$$

причем коэффициент $c$ пропорциональности (2) у взыскивающего подразделения вычисляется по его затратам $C_i$ и восстановленным годовым платежам $\sum_{n,RR\in Y_i} RR_n$ за прошлый год $i$ (параметры $Y_i$), коэффициент $c = max\left(C_1/\sum_{n,RR\in Y_1} RR_n, \dots, C_t/\sum_{n,RR\in Y_t} RR_n\right)$. Предполагается, что коэффициент $c < 100\%$, иначе требуется перестройка подразделения взыскания, поскольку затраты выше взыскиваемых сумм и такой исход мы опускаем. По имеющемся у автора материалам величина $c = 3 - 5\%$.

Такой метод расчета издержек отличается от методов, предлагаемых ранее, в которых $C_n$ есть самостоятельная сумма, а взыскание именуется как самостоятельное. Теперь издержки выбираются равномерно из всех восстановленных задолженностей $RR_n$ в едином процентном отношении, определяемом из многолетней практики взыскания, которую легко проверить. Взыскание является общим по сегменту LGD. Если кредит взыскивался, а результат равен нулю, то стоимость взыскание ложится на другие взысканные кредиты. Отсутствует понятие индивидуальной суммы, затраченной на конкретный кредит. Для таких издержек по ведению платежа формула (1) никогда не будет больше единицы. Тогда как (1) может быть меньше нуля, хотя очень редко, учитывая стоимость взыскания и то, что сумма взыскания не должна быть больше полной экспозиции к дефолту $\hat{E}_0$ с учетом дисконта. Авторы, учитывающие абсолютные издержки по взысканию (бухгалтерский, экономический подход [10]), имели значения LGD больше единицы. Например, в работах [11] значения больше единицы принимали 3-7% всех наблюдений соответственно.

В задачах учета сумм значение LGD>1 не проверяемо и не взыскиваемо, поэтому предлагается для всех моделей LGD учитывать затраты предложенным выше подходом и строгую установку $\mathrm{LGD} \leq 1$ использовать для дальнейших выкладок. Поскольку, эффективно получается в (1), что $\widehat{LGD}(0) = 1 - 1/\hat{E}_0 \sum_{n=1}^{N} \widehat{RR}_n / (1+r)^{\frac{n}{12}}$, где $\widehat{RR}_n = RR_n \cdot (1-c)$, то есть платежи $\widehat{RR}_n$ просто меньше на $c$ процентов относительно исходных. Поэтому мы далее учитываем совместно поступления и издержки как $\widehat{RR}_n$, поскольку пропорционально распределили издержки по поступившим суммам. В формулу (2) возможно заложить более сложный учет сумм взыскания на процесс восстановления, требующий большего числа параметров, однако в работе предлагается ограничиться простой пропорциональностью.

Если мы учитываем потери не с момента начала взыскания после дефолта $d$ , то LGD на текущую дату ID $t = \tau - d$ рассчитываются по формуле

$$\widehat{LGD}(t) = 1 - \frac{1}{\hat{E}(t)} \sum_{n=t+1}^{N} \frac{\widehat{RR}_n}{(1+r)^{\frac{n-t}{12}}} \qquad (3)$$

Полная экспозиция к дефолту $\hat{E}(t)$ есть просроченный основной долг, а также срочные и просроченные проценты, комиссии, штрафы, пени и иные платежи, начисленные по условиям Кредитного договора. Это извлекается из учетных информационных систем банка по состоянию на необходимую дату (это $\tau$, т.е. дата дефолта плюс $t$), а для учета стоимости взыскания применяется техника такая же, как и для $\widehat{LGD}(0)$ (1). Видно, что второе слагаемое (3) состоит из дисконтированных к дате дефолта восстановлений, которые затем наращиваются по методу сложных процентов (с учетом ставки $r$) к отчетному моменту $t$. Из (3) очевидно, что для не полностью взысканных ссуд при $t \to N$, $\widehat{LGD}(t) \to 100\%$, а при $t > N$, $\widehat{LGD}(t) = 100\%$. Максимальный горизонт взыскания N может не быть ограничен (математически $N \to \infty$).

Согласно изложенному выше, для модели $\widehat{LGD}_{ID}(0)$ требуется значение $\widehat{LGD}_{WD}$ , а математические граничные условия для фактических (наблюдаемых) LGD для каждого заемщика имеют вид

1. $\widehat{LGD}_{ID}(0) = \widehat{LGD}_{WD}$; (4)

2. $\lim\limits_{t\to\infty} \widehat{LGD}_{ID}(t) = 100\%$, где «∞» означает максимальный горизонт взыскания равный N. Модельное значение $\widehat{LGD}_{ID}(0)$ должно соответствовать статистике взысканий по сегменту дефолтов, на которых обучается модель $\sum_{j=1}^{All} \widehat{LGD}_{ID}(0)^j = \sum_{j=1}^{All}\left(1-\frac{1}{\hat{E}_0^j}\sum_{n=1}^{N}\frac{RR_n^j\cdot(1-c)}{(1+r^j)^{\frac{n}{12}}}\right)$.

Значение $\hat{E}(t)$ учитывает основной долг и неизменные проценты на любые даты t после дефолта (взыскание без реструктуризации), то эта сумма будет оцениваться как[1]

$$\hat{E}(t) = (1+r)^{\frac{t}{12}}\cdot\left(\hat{E}_0-\sum_{n=1}^{t}\frac{\widehat{RR}_n}{(1+r)^{\frac{n}{12}}}\right) \quad (5)$$

Т.е. из формулы (5) следует, что к отчетному моменту $t$ сумма восстановлений и затрат дисконтируется к дате дефолта, снижая сумму основного долга в дефолте $\hat{E}_0$, затем, на оставшуюся сумму проценты наращиваются к отчетному моменту $t$. Учитывая $\hat{E}_0 > 0$, объединяя (3), (5) и сокращая множитель $(1+r)^{\frac{t}{12}}$ получается оценка LGD ID в виде

$$\widehat{LGD}_{ID}(t) = \begin{cases} 0, & \text{если } \displaystyle\sum_{n=1}^{N}\frac{\widehat{RR}_n}{(1+r)^{\frac{n}{12}}} \geq \hat{E}_0 \\ \dfrac{\widehat{LGD}(0)}{1-R(t)}, & R(t) = \dfrac{1}{\hat{E}_0}\displaystyle\sum_{n=1}^{t}\frac{\widehat{RR}_n}{(1+r)^{\frac{n}{12}}} \end{cases} \quad (6)$$

В (6) явно отмечено, что оценка LGD ID располагается строго на отрезке [0,1]. Из (6) следует, что при постоянном и не переоцениваемом $\widehat{LGD}(0)$ значение $\widehat{LGD}(t)$ не убывает. А при конечном числе погашений график LGD ID выглядит как «ступеньки» на Рис.1.

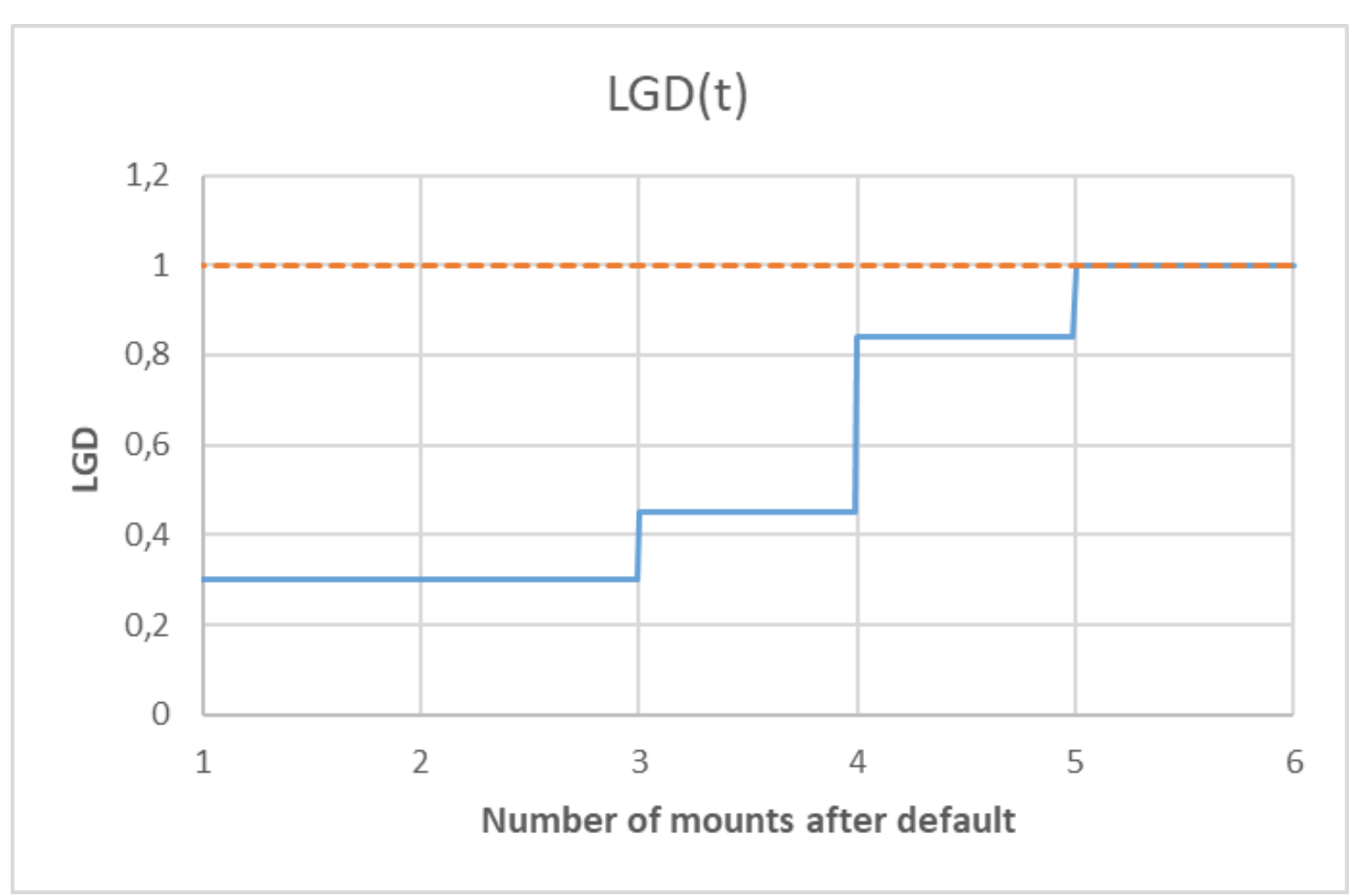

*Рисунок 1. Зависимость LGD(t) от срока t после дефолта*

В случае, крупной суммы (например, ипотека) Банк подает на заемщика в суд (tribunal) с требованием взыскания основного долга, доп. процентов и прочих издержек, накопившихся к моменту $t_{trbn}$ обращения, то после принятия обращения задолженность фиксируется. Получается, что по формуле (3) для $t > t_{trbn}$ проценты на возвращенные средства, дисконтированные к дате дефолта, продолжают начисляться, однако $E(t)$ наращенные на оставшиеся суммы проценты учитываться перестает. Это нарушает баланс активов и пассивов в сделке в промежуточных точках между взыскиваемыми суммами, поскольку экономически восстановительные платежи идут на погашение задолженности, на остаток которой начисление процентов останавливается с момента

---

[1] Следует из очевидного разностного уравнения: $E_t = (1+r)^{\frac{1}{12}}\cdot E_{t-1} - \widehat{RR}_t$

$t_{trbn}$. Это приводит к ложному падению $\widehat{LGD}(t)$ по (3) на участках «ступенек» после возврата заемщиком части КТ в отсутствие еще каких-либо фактических действий, поэтому такой разрыв требует установки ограничения справедливости формулы (3). А именно, формула (3) справедлива только в датах возврата любой величины долга. Иные точки справа следует учитывать только по точке зачисления ближайшей суммы возврата, которая находится слева, включая ноль. Это чисто математическое ограничение на уровне строгости, которое на результат действий не имеет, но указывает, что формула (3) не применяется для всех возможных дат либо считается константой по предыдущей дате платежа/объявления дефолта.

## 3. Модель восстановления ссуд после дефолта во времени

В основе модели динамики восстановления после дефолта – эмпирический двухпараметрический закон динамики восстановления. Он подтвержден на практике для большого числа портфелей дефолтов в разных сегментах КТ разных банков [12], [13]. Закон справедлив для суммы $\hat{R}(t) = \sum_{n=0}^{t} R_n$ такой, что

$$\hat{R}(t) \approx R(t) = R_\infty \cdot \left(1 - e^{-\frac{t}{T}}\right) \tag{7}$$

где $R_n = \frac{1}{All}\sum_{j=1}^{All} \frac{1}{\hat{E}_0^j} \frac{RR_n^j \cdot (1-c)}{(1+r^j)^{\frac{n}{12}}}$, $T$ – средний срок восстановления (мес.), $R_\infty \leq 100\%$ предельный уровень восстановления.

Для применения оценки на двух точках выписывается отношение объема $R^2$ возвратов во второй период $n \in (\tau, 2\tau]$ к объему $R^1$ за первый период $n \in (0, \tau]$, получается $\frac{R^2}{R^1} = \frac{e^{-\tau/T} - e^{-2\tau/T}}{1 - e^{-\tau/T}} = e^{-\tau/T}$, где $\tau$ задано, откуда можно дать простое двухточечное решение для $T = -\tau/ln\left({R^2}/{R^1}\right)$. Следует выбирать такое $\tau$, чтобы $2 \cdot \tau > T$, то есть достаточно ${R^2}/{R^1} \leq 0.6$. Соотношение ${R^2}/{R^1}$ вычисляется с погрешностью по исходным данным разных восстановлений $R^1 = \sum_{i=1}^{n_1} R_i^1$, $R^2 = \sum_{i=1}^{n_2} R_i^2$, где $R_i^1 = \frac{1}{\hat{E}_0^i}\sum_{k\in(0,\tau]} \frac{RR_k^i \cdot (1-c)}{(1+q)^{\frac{k}{12}}}$, $R_i^2 = \frac{1}{\hat{E}_0^i}\sum_{k\in(\tau,2\tau]} \frac{RR_k^i \cdot (1-c)}{(1+q)^{\frac{k}{12}}}$.

Опираясь на формулу стандартного отклонения для соотношения ${R^2}/{R^1}$ из стр. 81 [15], предполагая отсутствие корреляции между поступлениями первого и второго периодов, получаем соотношение и погрешность

$$\gamma = \frac{R^2}{R^1}, T = \frac{-\tau}{ln\,(\gamma)}$$

$$\sigma\gamma = \sigma_\gamma \cdot \gamma, \sigma_\gamma = \sqrt{\frac{\sum_{i=1}^{n_1}\left(R_i^1 - {1}/{n_1}\sum_{i=1}^{n_1} R_i^1\right)^2}{\left(\sum_{i=1}^{n_1} R_i^1\right)^2} + \frac{\sum_{i=1}^{n_2}\left(R_i^2 - {1}/{n_2}\sum_{i=1}^{n_2} R_i^2\right)^2}{\left(\sum_{i=1}^{n_2} R_i^2\right)^2}} \tag{8}$$

Решение (8) позволяет валидировать текущую зависимость $R_*(t)$ с функцией $e^{-\frac{t}{T_*}}$, для этого анализируется $t = \left|\frac{e^{-\tau/T_*} - \gamma}{\gamma}\right|$ по отношению к $\sigma_\gamma$.

Для задачи (7) можно выявить последовательность, максимум значения в которой будет еще ближе к идеальному решению. Пусть известны $\hat{R}_n \cong R(n) < R_\infty$ для $n = 1 \ldots N$, где $N$ – горизонт взыскания (мес.). Положим $R_\infty(\theta) = \hat{R}_N + \theta \cdot (1 - \hat{R}_N)$, где $0 < \theta < 1$, $\sum_{n=1}^{N} p_n = 1$, $T_n(\theta) = -n/ln\left(1 - \hat{R}_n/R_\infty(\theta)\right)$, $T(\theta) = \sum_{n=1}^{N} p_n \cdot T_n(\theta)$. Погрешность составляет $\sigma^2(\theta) = \sum_{n=1}^{N} p_n \cdot \left(T_n(\theta) - T(\theta)\right)^2$. Проводится оптимизация $\sigma^2(\theta)$, получается оптимальное решение

$$\theta^* = arg \min_{0<\theta<1} \sigma^2(\theta),\ T = T(\theta^*),\quad R_\infty = \hat{R}_N + \theta^* \cdot (1 - \hat{R}_N) \qquad (9)$$

для которого погрешность $\sigma T = \sqrt{\sigma^2(\theta^*)}$. Используются возможные крайние варианты вероятностей для (9): $p_n = 1/N$ учитывает равновероятное погашение, $p_n = 2(N-n)/N/(N+1)$ учитывает преимущество веса восстановлений за начальные периоды, $p_n = 2 \cdot \mathrm{n}/N/(N+1)$ учитывает преимущество веса восстановлений за конечные периоды. Для применения на практике выбирается наиболее консервативный вариант, т. е. имеющий минимальную оценку $R_\infty$.

Для сопоставительной оценки:

1. Строятся графики $\hat{R}(t)$и $R(t|T, R_\infty)$ на основе параметров (9), в оценке равновероятного погашения и оценивается MAD-погрешность

$$MAD = \frac{1}{N}\sum_{n=1}^{N}\left|\hat{R}(n) - R(n|T, R_\infty)\right| \qquad (10)$$

2. Оцениваются параметры распределения (9);
3. Оцениваются параметры для двухточечной оценки (8), погрешность и параметр $t$.

Рассматривается два примера для дефолтов по жилищному и потребительскому кредитованию, результаты даны в Таб. 1.

*Таблица 1. Оценки параметров восстановления*

| **Сегмент** | **Жилищное кредитование** | **Потребительское кредитование** |
|---|---|---|
| Дефолты, $M$, шт. | 2374 | 29500 |
| $\theta^*$ | 0,38 | 0,00139 |
| $R_N$ | 66,4% | 21,5% |
| $R_\infty(\theta^*)$ | 79,2% | 21,6% |
| $T(\theta^*)$, мес. | 30,82 | 11,32 |
| $\sigma T(\theta^*)$, мес. | 1,2 | 0,6 |
| MAD | 0,77% | 0,14% |
| Задано $\tau$ (год) | 2 | 1 |
| $\gamma = \frac{R^2}{R^1}$ | 0,4585 | 0,3490 |
| Двухточечное решение $T = -\tau/ln\left({}^{R^2}/_{R^1}\right)$ | 30,78 | 11,40 |
| Значение $t = \left\|\frac{e^{-\tau/T_*} - \gamma}{\gamma}\right\|$ | 0,1% | 0,75% |
| Погрешность $\sigma_\gamma$ | 3,91% | 2,1% |

И таблицы видно, что погрешность двухточечного решения гораздо больше, чем разница более точного определение $T(\theta^*)$ методом (9) и двухточечного методом (8). Это означает, что двухточечный метод дает адекватное решение и может применяться для определения среднего времени восстановления $T$. Отклонение экспоненциального решения (7) от кривой погашений демонстрируется параметром MAD, который ниже 1% и погрешностью $\sigma T(\theta^*)$, которая ниже 5% от значения $T$ в представленных примерах. Приведенные примеры констатируют высокий уровень сходимости для представленной зависимости и верный выбор средних.

## 4. Оценка LGD после дефолта

Пусть в любой из моментов времени после дефолта имеется оценка $\widetilde{LGD}(0) = L \leq 1$, которая справедлива для начального момента дефолта и распространяется на любое время после дефолта $t > 0$, тогда из (7) следует $\hat{R}(t) = (1 - L) \cdot \left(1 - e^{-\frac{t}{T}}\right)$. Значение LGD ID моделируется как вероятность потерь после дефолта, которая эквивалентна (6),

$$\widetilde{LGD}(t) = \begin{cases} \left(1 + \dfrac{1 - L}{L} \cdot e^{-\frac{t}{T}}\right)^{-1}, & L > 0 \\ 0, & \text{в остальных случаях} \end{cases} \qquad (11)$$

Очевидно, $\widetilde{LGD}(0) = L$. Асимптотически, при $L > 0$ из (11) следует $\widetilde{LGD}(t) = 1 - \varepsilon(t) + O\left(\varepsilon^2(t)\right)$, где $\varepsilon(t) = (1 - L)/L \cdot e^{-t/T}$. Очевидно, что для предложенного подхода построения модели LGD после дефолта выполнение граничных условий (4) обеспечено, поскольку $LGD(0)$ фиксировано, при этом $\widetilde{LGD}(t) \to 1$ при $t \to \infty$, а возрастание обеспечивается максимальным изменением значения $L$, которое меньше экспоненты от срока восстановления.

Значение $L$ может определяться двумя способами. Первый из них – это непрерывное определение значения $L = L(t)$ по величинам объектов в выбранном векторе детерминантов. Как правило должны учитываться суммы восстановления, решение дефолтеров о погашении, прозрачность и т. д. Для такого сценария требуется получение дополнительной модели, основанной на данных дефолтеров. Валидация такой модели должна основываться на наборе проверок и тестов как для метода LGD WD. Второй способ – это построение модели по данным о величине LGD WD и текущему значению поступлений. Вторым способом займемся далее.

## 5. Байесовская модель оценки уровня восстановления

В основе модели взыскания лежит допущение, что среднее восстановление для m-го КТ подчиняется уравнению (7) с постоянным параметром $T$ и индивидуальным для m-го КТ параметра восстановления на бесконечном времени $R_\infty = 1 - L$, который в рамках предлагаемого подхода подлежит апостериорному уточнению. Полагается, что оценен параметр $T$ среднего срока восстановления (7) на уровне обучающей выборки. При интегрировании (8) по аргументу $\tau \in [0, t]$ получается тождество $\hat{R}_\infty(t) = \left(T \cdot \hat{R}(t) + \int_0^T \hat{R}(\tau) d\tau\right)/t = \left(T \cdot \hat{R}(t) + t \cdot \hat{R}(t) - \int_0^T \hat{R}'(\tau) \tau d\tau\right)/t.$ Учитывая (8), а также $\hat{R}'(t) = RR_t$ и $\hat{R}(t) = \sum_{n=0}^{t} RR_n$, получается апостериорная оценка

$$\hat{R}_\infty(t) = \sum_{n=0}^{t} RR_n + \frac{1}{t} \sum_{n=0}^{t} RR_n \cdot (T - n) \qquad (12)$$

**Лемма**. *Оценка (12) является состоятельной*[2].

Доказательство леммы представлено в Приложении. Пусть дефолтный заемщик в месячные периоды $k = 0,1, \dots\ t_k < t$ обеспечил дисконтированные восстановления $RR_{t_k}$. Наступивший срок $t$ после дефолта заемщика $d$ соответствует поколению $\tau = d + t$ с начала наблюдения $\tau = 0$, установлен средний срок восстановления $T$ (9). Будем предполагать, что $T$ является постоянным. Формула (12) дает реализуемый на практике алгоритм оценки апостериорного (по результатам наблюдений) наблюдения за заемщиком $L(t) = 1 - \hat{R}_\infty(t)$, который перепишется в виде:

$$\hat{R}_\infty(t) = \sum_{t_k \le t} RR_{t_k} + \frac{1}{t} \cdot \sum_{t_k \le t} RR_{t_k} \cdot (T - t_k), k = 1,2, \dots \\ RR_{t_k} = \frac{\widehat{RR}_{t_k}/\hat{E}_0}{(1+r)^{\frac{t_k}{12}}},\ t_0 = 1 \tag{13}$$

Очевидно, что оценка (13) будет существенно волатильна на ранних стадиях $t$ погашений (или ожидания таковых) дефолтного КТ, поэтому требуется использовать априорные WD-оценки восстановления из начала дефолта. Априорной оценкой уровня погашения $R_{\infty WD}$ может быть модельная оценка $R_{\infty WD} = 1 - LGD_{WD}$.

С позиции Байесовского подхода к оценке апостериорного распределения случайной величины при наличии ограниченного количества измерений n, разумно ожидать, что средние значения апостериорного распределения $\tilde{\mu}_n$ будет декомпозироваться по формуле[3]

$$\tilde{\mu}_n = w_n \cdot \mu_0 + (1 - w_n) \cdot \mu_n \tag{14}$$

где $\mu_0$ матожидание априорного распределения, $w_n \in (0,1]$ - вес, такой, что $w_0 = 1, w_n \to 0$ , при $n \to \infty$ , а $\mu_n$ – апостериорная оценка по результатам наблюдений восстановления $\hat{R}_\infty(t)$ (13). Значения $w_n, \mu_0, \mu_n$ дает первая строка в таблице Приложения в виде $\frac{1}{1+t/T}$, $1 - LGD_{WD}^m$, $\hat{R}_\infty^m(t)$ соответственно, значение $T$ определяется из оценки (9). Тогда полная модель апостериорного значения $\tilde{R}_\infty$ для ссуды в дефолте строится по Байесовской формуле

$$\tilde{R}_\infty(t) = \frac{1}{1 + t/T} \cdot (1 - LGD_{WD}) + \frac{t/T}{1 + t/T} \cdot \hat{R}_\infty(t) \tag{15}$$

Окончательно применяется формула (11) для получения результата, при $L = L(t) = 1 - \tilde{R}_\infty(t)$.

Предложенный байесовский сценарий не требует построения отдельной модели LGD ID на основе детерминантов. Для него достаточно построения удачной консервативной модели LGD WD, расчета базового срока восстановления по сегменту и переоценки предельного уровня восстановления $\hat{R}_\infty(t)$ на текущий отчетный период $t$ после дефолта на основе наблюдаемого ряда погашений.

Для практического примера рассмотрим результат применения прогноза к метрикам дефолтных кредитов. В качестве первого – рассмотрим десятилетний опыт взыскания по ипотеке (mort) одного крупного банка с количеством случаев $N_{mort} > 6000$, в качестве второго примера – десятилетний опыт взыскания по потребительскому кредитованию (cl) с количеством случаев $N_{cl} > 200\ 000$. Данные по каждому кредиту: объем ссуд взыскания, ставка по ссуде, объемы погашения в каждый месяц после дефолта на протяжении 124 месяцев (предельный срок). Таким образом мы рассчитываем приведенные погашения (13) и объем погашения для каждой ссуды, сокращая к нулю, если дельта он будет больше 100%. В качестве среднего объема погашения возьмем среднее по объёму уровень погашения, для ипотеки $RR_{mort} = 28\%$, для потребительского кредита получается $RR_{cl} = 13.2\%$. Эти данные мы используем в качестве основных, не учитывая, например, другие виды погашения (по ипотеке - через продажу жилья). Затраты на взыскание учитывать не будем, их учет не даст принципиального изменения результатов, поскольку он просто повышает кредитную нагрузку до 5%, не приводя к расхождению в расчетах. Точные значения $LGD(t)$ после дефолта рассчитываются по формуле (6), зная все восстановления в моменты после дефолта. Прогнозные значения после дефолта $LGD(t)$ и восстановлений $RR(t)$ для каждой ссуды рассчитывается по (13), (15), где в качестве значений $T$ берутся модели восстановления ссуд по (9). В качестве опоры расчетов используются также средние значения восстановлений по формуле $RR_j(t) = RR_j \cdot \left(1 - e^{-t/T_j}\right)$, $j = mort, cl$. Мы используем среднее значение численного измерения по ссуде в виде

$$\langle r \rangle = \frac{\sum_{i=1}^{N} r_i \cdot \hat{E}_0^i}{\sum_{i=1}^{N} \hat{E}_0^i} \tag{16}$$

применяя его также к $LGD(t)$ и далее.

---

[2] Состоятельная оценка в математической статистике — это точечная оценка, сходящаяся по вероятности к оцениваемому параметру.

[3] Для семи наиболее известных сопряженных априорных распределений декомпозиции средних значений представлены в Приложении.

На Рис. 4 поставим три значения $\langle LGD(t) \rangle$ на ось времени после дефолта — это точное, прогнозное и среднее, как представлено выше. Первый рисунок – это ипотека, второе – потребительский кредит.

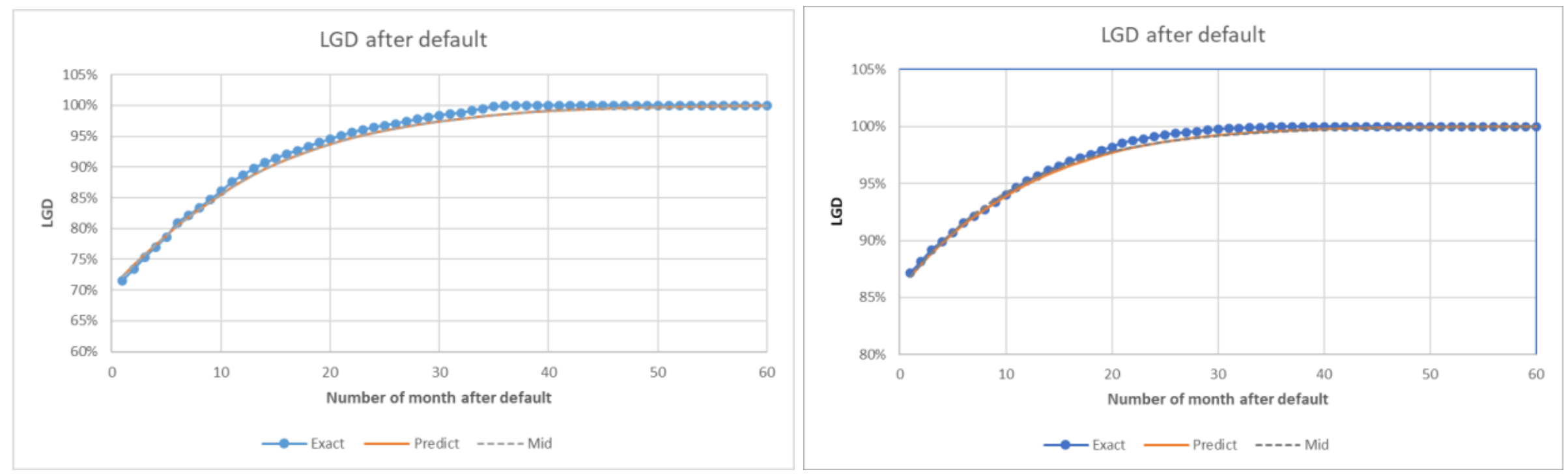

*Рисунок 2. Значение $LGD(t)$ для расчета по значениям восстановлений (синий), а также для прогнозный значений (коричневый) и средних значений (серый). Первый рисунок – ипотека (mort), второй рисунок – потребительское кредитование (cl).*

Поведение рассчитанных точных последовательностей LGD аналогично [14] расчетам ELBE для метрик нескольких стран. Однако видно, что расчетное значение $LGD(t)$ очень близко между точным и рассчитанным по прогнозу. Этот практический расчет почти не дает разницу между прогнозной и средней значениями.

Чтобы показать, что прогнозное значение существенно лучше среднего, мы возьмем две метрики на основании восстановлений: первое – это разница значений $dR_1 = \langle \left| \tilde{R}(t) - R(t) \right| \rangle$ (Predict-Exact), а второе $dR_2 = \langle |RR(t) - R(t)| \rangle$ (Md-Exact). Здесь прогноз $\tilde{R}(t)$ из формулы (15), а $R(t)$ точное значение (6), где $RR(t)$ усредненное значение (16), которое берется без учета восстановлений. На Рис.5 показано как ведут себя метрики $dR_1, dR_2$ для двух портфелей (первый рисунок – ипотека, второй – потребительское кредитование).

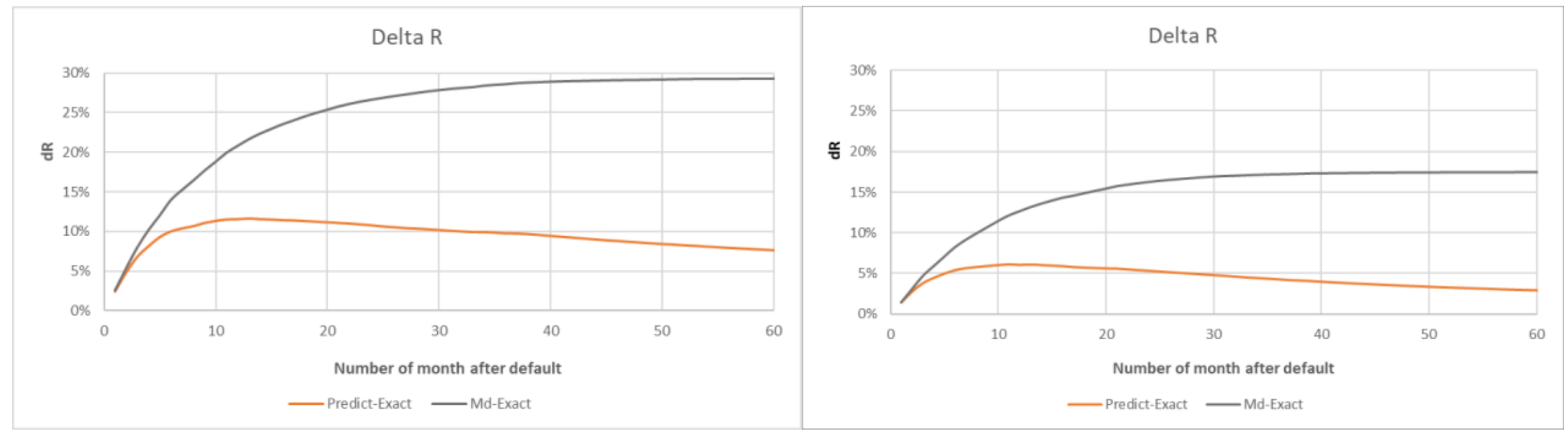

*Рисунок 3. Значение усредненных метрик модуля разницы восстановлений прогнозного/точного (коричневый) и среднего/точного (серый). Первый рисунок – ипотека, второй рисунок – потребительское кредитование*

По рисункам видно, что метрика разницы с точным от средней сильно отстает от метрики разницы с точным от прогнозной, поэтому представленную прогнозную метрику имеет прямой смысл вставлять в модель. По Рис. 2 между прогнозной и средней разницы почти нет, в рамках усредненной метрики $LGD(t)$ портфеля, однако по индивидуальным показателям разница существенная Рис. 3. Совокупность рисунков показывает работоспособность предложенной байесовской модели. Возможно, модель (12) удастся расширить, чтобы получить более близкие к точным данные на выходе, но мы остановимся на том, что предложили.

## 6. Заключение

В предложенной работе показано как определять LGD ID с подготовкой или без подготовки отдельной модели. Для этого требуются:

1. Подготовленная для разных сроков после дефолта модель $\mathrm{LGD}(0)$ или модель $\mathrm{LGD\ WD}$;
2. Расчет среднего времени для экспоненциального погашения ссуд после дефолта;
3. Размеры параметров в выбранном векторе детерминантов или значения объемов и моментов погашений после дефолта вместе со ставкой кредитования по каждой ссуде.

Для модели LGD ID требуется соблюдение граничных условий (4). Значение среднего периода для погашения ссуд после дефолта рассчитывается для искомого сегмента по формулам (8) или (9). Проведено исследование на портфелях дефолтов, достоверно показаны сходимости для среднего времени экспоненциального погашения.

Для методики с использованием обновляемой модели $\mathrm{LGD}(0)$ после дефолта требуется формула (11) для расчета итогового $\widetilde{LGD}(t)$. Если учитывается только модель LGD WD, то постоянно (в каждый момент $t$) определяется уровень восстановления на не ограниченном времени после дефолта. Для этого используется прогноз по модели восстановления после дефолта (13), затем с помощью метода оценки апостериорного восстановления (15) определяется искомое $\widetilde{LGD}(t)$ по формуле (11). При применении апостериорного распределения в практических примерах получены кривые восстановлений, практически совпадающее с точным. При оценке сходимости на индивидуальных уровнях показаны существенные улучшения для апостериорного распределения (15).

Подготовка оценки опиралась на модель общего взыскания на восстановление по сегменту LGD, по типу (2). Только такой тип взыскания позволяет установить $\mathrm{LGD} \leq 1$, а это установка, требуемая для выкладок оценки $\widetilde{LGD}(t)$. Коэффициент взыскания валидируется по документам Службы взыскания.

## ИСТОЧНИКИ

## ПРИЛОЖЕНИЕ

**Доказательство Леммы.**

Требуется доказать, что $\widehat{R}_N = \sum_{n=0}^{N} RR_n + \frac{1}{N}\sum_{n=0}^{N} RR_n \cdot (T-n) \to R_\infty = \lim_{N\to\infty} \sum_{n=0}^{N} RR_n$.

Имеем, $RR_n \geq 0$, $\widehat{R}_N = \left(1+\frac{T}{N}\right) \cdot \sum_{n=0}^{N} RR_n - \sum_{n=0}^{N} RR_n \cdot \frac{n}{N}$. Для доказательства того, что $\widehat{R}_N \to R_\infty$ при $N \to \infty$ достаточно показать, что $\sum_{n=0}^{N} RR_n \cdot \frac{n}{N} = S_N \to 0$.

Положим $n_*(N) = N^{\frac{1}{1+\gamma}}$ для некоторого $\gamma > 0$, $S_N = P_N + Q_N$, где $P_N = \sum_{n=0}^{n_*(N)} RR_n \cdot \frac{n}{N}$, $Q_N = \sum_{n=n_*(N)+1}^{N} RR_n \cdot \frac{n}{N}$. Очевидно, $P_N < \sum_{n=0}^{n_*(N)} RR_n \cdot \frac{n_*(N)}{N} = N^{-\frac{\gamma}{1+\gamma}} \cdot \sum_{n=0}^{n_*(N)} RR_n \leq \frac{R_\infty}{N^{\frac{\gamma}{1+\gamma}}} \to 0$ при $N \to \infty$. Второе слагаемое, $Q_N \leq \sum_{n=n_*(N)+1}^{N} RR_n \leq \sum_{n=n_*(N)+1}^{\infty} RR_n \to 0$, поскольку ряд $\sum_{n=0}^{\infty} RR_n$ сходится, а нижняя граница $n_*(N)+1 \to \infty$ при $N \to \infty$. Значит доказано, что $S_N \to 0$.

**Декомпозиции средних значений некоторых наиболее известных сопряженных апостериорных распределений**

В байесовском подходе оптимальным образом используется информация из двух источников:

- априорная информация о моделируемом объекте (информация, полученная из предыдущих исследований или теоретических предположений)
- сстатистическая информация, содержащаяся в результатах наблюдений.

Обновленная информация (апостериорная вероятность) - результат применения формулы Байеса:

$$p(\theta|Y) = \frac{p(\theta) \cdot L(Y|\theta)}{\int L(Y|\theta) p(\theta) dp(\theta)}$$

Где $Y$- вектор реализации случайных значений, $p(\theta)$ – плотность априорного распределения, $L(Y|\theta)$ – функция правдоподобия, как плотность вероятности реализаций $Y$ при заданном параметре $\theta$, $\int L(Y|\theta)p(\theta)dp(\theta)$ - полная вероятность, выполняющая роль нормирующего множителя и не зависящая от вектора параметров $\theta$.

Искомая $p(\theta|Y)$ - апостериорная функция плотности вероятности, включающая как априорную (через априорную плотность распределения вектора параметров), так и (через функцию правдоподобия) апостериорную информацию. Полученную апостериорную функцию плотности вероятности можно охарактеризовать мерами центральной тенденции (математическим ожиданием или модой), дисперсии и скошенности.

Общая логическая схема байесовского оценивания значения параметров распределения представлена на следующем рисунке

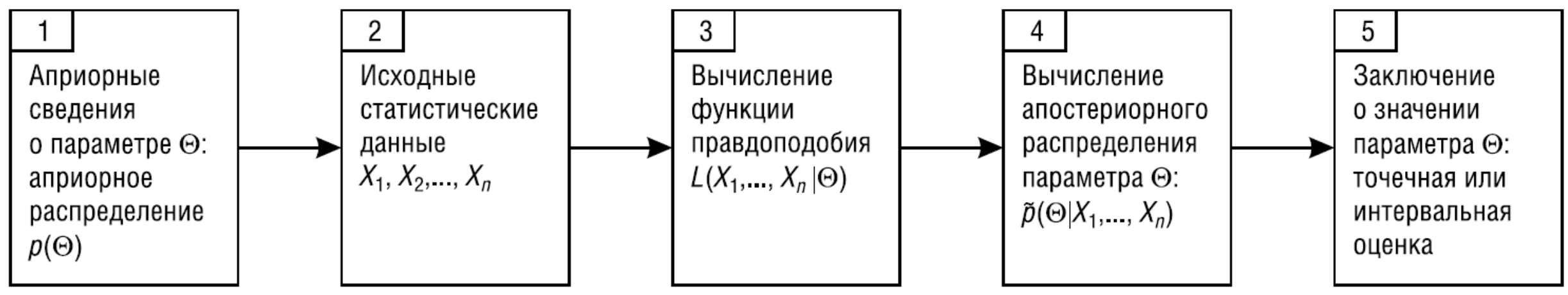

*Рисунок. Общая логическая схема байесовского подхода*

Если апостериорное распределение $p(\theta|Y)$ принадлежит тому же семейству вероятностных распределений, что и априорное распределение $p(\theta)$ (т.е. имеет тот же вид, но с другими

параметрами), то это семейство распределений называется ***сопряжённым*** семейству функций правдоподобия $L(Y|\theta)$. ***При этом распределение $\boldsymbol{p(\theta)}$ называется сопряжённым априорным распределением к семейству функций правдоподобия $\boldsymbol{L(Y|\theta)}$***. Семейство сопряженных распределений не многочисленно. В Таблице представлены некоторые известные сопряженные распределения из (Aivazian, 2008), выданы соответствующие значения $\boldsymbol{w_n, \mu_0, \mu_n}$ для средних значений апостериорного распределения $\boldsymbol{\hat{\mu}_n = w_n \cdot \mu_0 + (1 - w_n) \cdot \mu_n}$.

*Таблица. Семейство законов распределения вероятностей (ЗРВ), сопряженные априорные и апостериорные распределения, а также параметры декомпозиции средних значений* $\bar{x} = \frac{1}{n}\sum_{i=1}^{n} x_i$, $\bar{g} = \left(\prod_{i=1}^{n} x_i\right)^{\frac{1}{n}}$.

| № | LPD of the observed population | Conjugate apriority LPD $\boldsymbol{p(\theta)}$, expressions for $\boldsymbol{E\theta}$ и $\mathbf{D}\boldsymbol{\theta}$ | Posterior LPD $\boldsymbol{p(\theta \mid x_1, x_2, \dots x_n)}$ and expressions for its parameters | Decomposition parameters of medium values: $\boldsymbol{w_n, \mu_0, \mu_n}$ for posterior distribution |
|---|---|---|---|---|
| 1 | $(\theta; \sigma^2)$ – normal,<br>$f(x \mid \theta) = \frac{1}{\sqrt{2\pi}\sigma} e^{-\frac{(x-\theta)^2}{2\sigma^2}}$<br>*(the variance value $\sigma^2$ is known)* | $(\theta_0; {\sigma_0}^2)$ – normal,<br>$E\theta = \theta_0; D\theta = {\sigma_0}^2$<br>($\theta_0$ and ${\sigma_0}^2$ are set) | $(\theta_0'; \sigma_0'^2)$ – normal,<br>where $\theta_0' = \frac{\bar{x}+\gamma\theta_0}{1+\gamma}$,<br>$\sigma_0'^2 = \frac{\sigma^2}{n(1+\gamma)}$ and<br>$\gamma = \frac{\sigma^2}{n{\sigma_0}^2}$ | $w_n = \frac{1}{1+1/\gamma}$<br>$\mu_0 = \theta_0$<br>$\mu_n = \bar{x}$ |
| 2 | Exponential<br>$f(x \mid \theta) = \begin{cases} \theta e^{-\theta x}, & x \geq 0 \\ 0, & x < 0 \end{cases}$ | $p(\theta) = \frac{\beta^\alpha}{\Gamma(\alpha)} \theta^{\alpha-1} e^{-\beta\theta}$, $(\theta > 0)$<br>are gamma distribution;<br>$E\theta = \frac{\alpha}{\beta}; D\theta = \frac{\alpha}{\beta^2}$<br>( $\alpha, \beta$ are set) | Gamma distribution with parameters<br>$\alpha' = \alpha + n$<br>$\beta' = \beta + n \cdot \bar{x}$ | $w_n = \frac{1}{1+n\frac{\bar{x}}{\beta}}$<br>$\mu_0 = \frac{\alpha}{\beta}$<br>$\mu_n = \frac{1}{\bar{x}}$ |
| 3 | $[0; \theta]$ are uniform distribution<br>$f(x \mid \theta) = \begin{cases} \frac{1}{\theta}, 0 \leq x \leq \theta \\ 0, x \notin [0; \theta] \end{cases}$ | $p(\theta) = \begin{cases} \frac{\alpha\theta_0^\alpha}{\theta^{\alpha+1}}, x \geq 0; \\ 0, x < 0 \end{cases}$<br>$(\theta > 0)$ are Pareto distribution;<br>$E\theta = \frac{\alpha\theta_0}{\alpha-1}; D\theta = \frac{\alpha{\theta_0}^2}{(\alpha-1)^2(\alpha-2)}$<br>*($\alpha, \theta_0$ are set)* | Pareto distribution with parameters<br>$\alpha' = \alpha + n$<br>$\theta_0' =$<br>$\max\{\theta_0, x_1, x_2, \dots,, x_n\}$ | $w_n = \frac{1}{1+n\cdot\frac{1}{\alpha-1}} \cdot \frac{1}{1-\alpha\frac{\theta_0'-\theta_0}{\theta_0'}}$<br>$\mu_0 = \frac{\alpha\theta_0}{\alpha-1}$<br>$\mu_n = \theta_0'$ |
| 4 | Poisson distribution<br>$P\{\xi = x\} = \frac{\theta^x}{x!} e^{-\theta}$,<br>$x = 0,1,2, \dots$ | $p(\theta) = \frac{\beta^\alpha}{\Gamma(\alpha)} \theta^{\alpha-1} e^{-\beta\theta}$, $(\theta > 0)$<br>are Gamma distribution;<br>$E\theta = \frac{\alpha}{\beta}; D\theta = \frac{\alpha}{\beta^2}$<br>( $\alpha, \beta$ are set) | Gamma distribution with parameters<br>$\alpha' = \alpha + n \cdot \bar{x}$<br>$\beta' = \beta + n$ | $w_n = \frac{1}{1 + n \cdot \frac{1}{\beta}}$<br>$\mu_0 = \frac{\alpha}{\beta}$<br>$\mu_n = \bar{x}$ |
| 5 | The binomial distribution<br>$P\{\xi = x\} = C_N^x \theta^x (1-\theta)^{N-x}$<br><br>(parameter value N is known) | $p(\theta) = \frac{\Gamma(a+b)}{\Gamma(a)\Gamma(b)} \theta^{a-1} (1-\theta)^{b-1}$<br>$(0 \leq \theta \leq 1)$ are Betta distribution;<br>$E\theta = \frac{a}{a+b}; D\theta = \frac{ab}{(a+b)^2(a+b+1)}$<br>( $a, b$ are set) | Betta distribution with parameters<br>$a' = a + n \cdot \bar{x}$<br>$b' = b + nN - n \cdot \bar{x}$ | $w_n = \frac{1}{1+n\cdot\frac{N}{a+b}}$<br>$\mu_0 = \frac{a}{a+b}$<br>$\mu_n = \frac{\bar{x}}{N}$ |
| 6 | Negative binomial distribution<br>$P\{\xi = x\} = C_{x-1}^{k-1} \theta^k (1-\theta)^{x-k}$<br>(parameter value k is known)<br>$x = k, k+1, \dots$ | $p(\theta) = \frac{\Gamma(a+b)}{\Gamma(a)\Gamma(b)} \theta^{a-1} (1-\theta)^{b-1}$<br>$(0 \leq \theta \leq 1)$ are Betta distribution;<br>$E\theta = \frac{a}{a+b}; D\theta = \frac{ab}{(a+b)^2(a+b+1)}$<br>( $a, b$ are set) | Betta distribution with parameters<br>$a' = a + kn$<br>$b' = b + n \cdot \bar{x} - kn$ | $w_n = \frac{1}{1+n\cdot\frac{\bar{x}-k}{a+b}}$<br>$\mu_0 = \frac{a}{a+b}$<br>$\mu_n = \frac{k}{\bar{x}-k}$ |
| 7 | Pareto distribution<br>$f(x \mid \theta) = \begin{cases} \frac{\theta x_0^\theta}{x^{\theta+1}}, x \geq x_0 \\ 0, x < x_0 \end{cases}$<br>(parameter $x_0$ is known) | $p(\theta) = \frac{\beta^\alpha}{\Gamma(\alpha)} \theta^{\alpha-1} e^{-\beta\theta}$, $(\theta > 0)$<br>are Gamma distribution;<br>$E\theta = \frac{\alpha}{\beta}; D\theta = \frac{\alpha}{\beta^2}$<br>( $\alpha, \beta$ are set) | Gamma distribution with parameters<br>$\alpha' = \alpha + n$<br>$\beta' = \beta + n \cdot \ln\left(\frac{\bar{g}}{x_0}\right)$ | $w_n = \frac{1}{1+n\cdot\frac{ln(\bar{g}/x_0)}{\beta}}$<br>$\mu_0 = \frac{\alpha}{\beta}$<br>$\mu_n = \frac{1}{ln(\bar{g}/x_0)}$ |